\documentclass{epl}

\title{Two mechanisms of pseudogap formation in Bi-2201:
Evidence from the $c$-axis magnetoresistance}
\shorttitle{Two mechanisms of pseudogap formation in Bi-2201}
\author{A. N. Lavrov \and Yoichi Ando \and S. Ono}
\shortauthor{A. N. Lavrov \etal}
\institute{Central Research Institute of Electric Power Industry, 2-11-1
Iwato-kita, Komae, Tokyo 201-8511, Japan}

\begin{document}

\maketitle

\begin{abstract}
Measurements of the $c$-axis resistivity and magnetoresistance have been
used to investigate the pseudogap (PG) behavior in
Bi$_{2+z}$Sr$_{2-x-z}$La$_x$CuO$_y$ (Bi-2201) crystals at various hole
densities. While the PG opening temperature $T^*$ increases with
decreasing hole doping, the magnetic-field sensitivity of the PG is
found to have a very different trend: it appears at {\it lower}
temperatures in more underdoped samples and vanishes in
non-superconducting samples. These data suggest that besides the
field-insensitive pseudogap emerging at $T^*$, a distinct one is formed
above $T_c$ as a precursor to superconductivity.
\end{abstract}

\section{Introduction}
In high-$T_c$ cuprates, the electronic density of states (DOS) near the
Fermi energy has been demonstrated to decrease gradually with decreasing
temperature, resulting in the pseudogap (PG) formation
\cite{review,STM,ARPES3,ARPES4}. This PG, which has been found to
progressively ``destroy'' the Fermi surface (FS) \cite{ARPES4,ARPES1},
is on one hand a challenge to the conventional view of the FS itself,
while on the other hand it allows one to reconcile the small number of
carriers that participate in the charge transport in underdoped cuprates
with the large FS observed by photoemission \cite{review,ARPES1}.

Although the existence of the PG has been documented by many experiments
\cite{review}, its nature and, particularly, its relation to the
superconductivity (SC) remain far from being clear. For example,
according to photoemission and surface-tunneling studies
\cite{STM,ARPES3,ARPES4}, the PG evolves smoothly into the SC gap below
$T_c$, which implies a precursor-pairing origin of the PG; in contrast,
recent observations of a distinct SC gap that coexists with the PG below
$T_c$ and tends to close at $T_c$ \cite{2Gap1,2Gap2} suggest that the PG
might have nothing to do with superconductivity \cite{Tallon,Tallon2}.
Apparently, a crucial test for the origin of the PG would be its
sensitivity to the magnetic field; however, studies of the
magnetic-field dependence of the PG have reported surprisingly
controversial results \cite{NMR1,NMR3,NMR2,overdoped,cMR}.

While the electronic DOS in cuprates has mostly been studied by
photoemission \cite{ARPES3,ARPES4,ARPES1} or surface-tunneling
\cite{STM,Rc_gap} spectroscopies, the extremely anisotropic nature of
the Bi-based cuprates, where the crystal structure itself forms a stack
of tunnel junctions, offers a possibility of {\it intrinsic} tunneling
spectroscopy \cite{2Gap1,2Gap2,tunnel}. In fact, it has been shown that
the $c$-axis transport in the Bi-based cuprates is governed by the
tunneling between CuO$_2$ planes and, accordingly, the $c$-axis
voltage-current characteristics directly reflect the DOS structure of
the CuO$_2$ planes \cite{2Gap1,2Gap2,tunnel}. It was reported that the
$c$-axis differential resistivity $\rho_c^d$ at high bias shows
virtually no temperature dependence, indicating that the tunneling
mechanism as well as the DOS away from the Fermi level are temperature
independent \cite{2Gap1}. However, the low-bias $\rho_c^d$ reveals a
pronounced upturn upon decreasing temperature below $T^*$, demonstrating
the opening of the PG near the Fermi energy \cite{2Gap1,2Gap2,tunnel}.
This tunneling behavior of $\rho_c^d$ is in sharp contrast to the
metallic in-plane resistivity that is governed by scattering and is
believed to decrease with the PG opening \cite{Ra}.

Since the conventional $c$-axis resistivity $\rho_c$ corresponds to
$\rho_c^d$ at zero bias, $\rho_c$ of the Bi-based cuprates measured
above $T_c$ reflects the DOS at the Fermi energy \cite{2Gap1,2Gap2}.
This means that $\rho_c$ provides a convenient probe to trace the
behavior of the PG, whatever its nature is. Indeed, the characteristic
temperature $T^*$ below which $\rho_c(T)$ shows a ``semiconducting''
upturn coincides well with the pseudogap temperatures determined by
other methods \cite{Rc_gap}. Correspondingly, the $c$-axis
magnetoresistance (MR), $\Delta \rho_{c}/\rho_{c}$, is expected to bear
information on the magnetic-field dependence of the PG, and the negative
$c$-axis MR \cite{cMR,Bi_MR} has been proposed to reflect a partial
recovery of the DOS with magnetic field. So far, however, most of the MR
studies were done on cuprates at nearly optimum doping, where $T^*$
closely approached $T_c$, and thus it was difficult to distinguish
whether the observed effect was related to the SC fluctuations
\cite{Bi_MR,Varlamov} or to the normal-state PG \cite{cMR}. Therefore,
it is desirable to extend the MR study to the heavily-underdoped region
where $T^*$ and $T_c$ are far apart.

In this Letter, we report the study of $\rho_c$ and the $c$-axis MR in
Bi$_2$Sr$_{2-x}$La$_x$CuO$_y$ (BSLCO) and Bi$_{2+z}$Sr$_{2-z}$CuO$_y$
(BSCO) single crystals for a wide doping range, from superconducting
compositions with $T_c$ of $30\un{K}$ to heavily-underdoped
non-superconducting ones. The evolution of $\rho_c(T)$ indicates that
$T^*$ increases with decreasing hole doping without being interrupted by
the disappearance of SC. In heavily-underdoped samples, the MR appears
to be extremely small in a wide temperature range below $T^*$,
suggesting a magnetic-field-insensitive nature of the normal-state PG.
However, it is found at the same time that a noticeable negative MR,
indicating a recovery of the pseudogapped DOS, starts to show up upon
approaching $T_c$. We discuss that the appearance of the negative MR
well below $T^*$ is likely to be associated with a secondary pseudogap,
which can be suppressed with magnetic fields. The data therefore point
to a possibility of two distinct mechanisms for the PG in cuprates: The
first one is insensitive to the magnetic-field, gains strength with
decreasing hole doping, and ultimately causes a transition into an
insulating state; the second one is sensitive to the magnetic field and
is clearly related to the superconductivity.

\section{Experimental methods}

Single crystals with nominal compositions of
Bi$_2$Sr$_{2-x}$La$_x$CuO$_y$ ($x=0.2-1.0$) and
Bi$_{2.2}$Sr$_{1.8}$CuO$_y$ are grown by the floating-zone method
\cite{growth}. Substituting Sr with La reduces the hole density in
Bi$_2$Sr$_{2-x}$La$_x$CuO$_y$; optimum doping corresponds to $x
\approx0.45$ ($T_c\approx30\un{K}$) and the system becomes
non-SC for $x > 0.9$ (heavily underdoped). The
Bi$_{2.2}$Sr$_{1.8}$CuO$_y$ crystal is nearly optimally doped (its hole
density can be estimated \cite{Hanaki} to be $\sim$17\% per Cu), and yet
its $T_c$ (onset at $\sim 3\un{K}$) is much smaller than in slightly
underdoped BSLCO with $x=0.5$ ($T_c=29.7\un{K}$; $\Delta T_c\approx
0.35\un{K}$), whose hole density is $\sim$15\% per Cu \cite{Hanaki}.

The $\rho_c$ measurements are done on samples with typical sizes of
$1\times 1\times 0.1\un{mm^3}$ using the ac four-probe technique. To
provide a homogeneous current flow along the $c$-axis, two current
contacts are painted to almost completely cover the opposing $ab$-faces
of the crystal; two voltage contacts are placed in the small windows
reserved in the center of the current contacts \cite{cont}. The MR
measurements are carried out by sweeping the magnetic field to
$\pm14\un{T}$ at fixed temperatures stabilized by a capacitance sensor
with an accuracy of $\sim1\un{mK}$; this method allows reliable
measurements of $\Delta\rho_c/\rho_c$ as small as $10^{-5}$ at
$10\un{T}$.

\section{Results and discussion}

\begin{figure}
\onefigure[width=10.4cm]{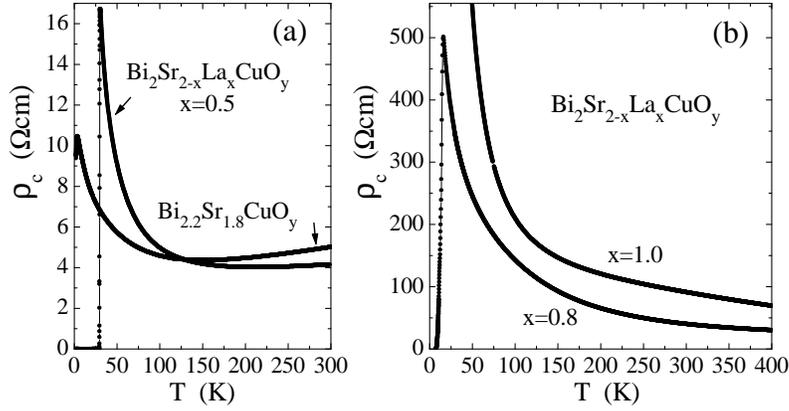}
\caption{The $c$-axis resistivity of nearly optimally doped (a) and
underdoped (b) BSLCO and BSCO single crystals. The data show the
``semiconducting'' low-temperature upturn, which marks the pseudogap
opening.}
\label{fig1}
\end{figure}

All the Bi-2201 samples demonstrate a steep increase in $\rho_c(T)$ at
low temperatures (fig.~\ref{fig1}), which has been shown to originate
from the PG formation and corresponding decrease in the DOS at the Fermi
energy \cite{2Gap1,2Gap2,Rc_gap,tunnel}. One can infer that the onset of
the $\rho_{c}$ upturn shifts to higher temperatures with decreasing hole
doping, as has been reported for other cuprates
\cite{review,ARPES3,Tallon,Rc_gap}, and that the evolution of the
$\rho_c(T)$ behavior is apparently uninterrupted by the disappearance of
SC [fig.~\ref{fig1}(b)].

\begin{figure}[!b]
\onefigure[width=10.6cm]{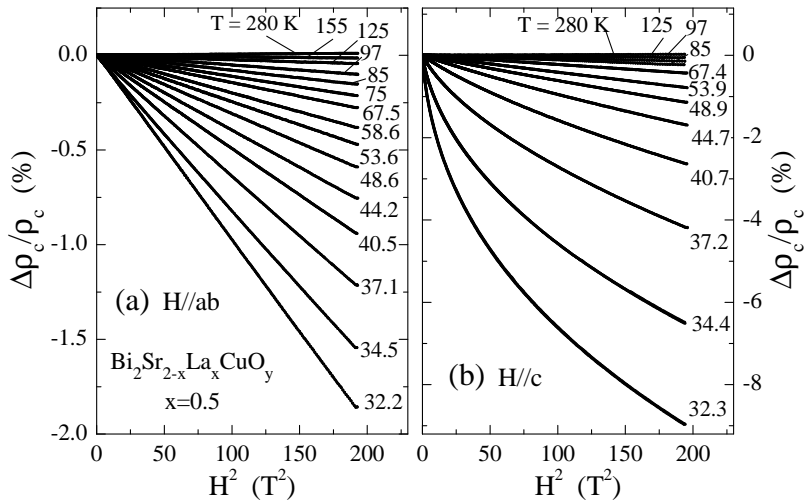}
\vspace{-5pt}
\caption{$H$ dependences of the transverse (a) and longitudinal (b)
$c$-axis MR for BSLCO with $x=0.5$. The field is swept between $\pm
14\un{T}$.}
\label{fig2}
\end{figure}

\begin{figure}[t]
\onefigure[width=11.8cm]{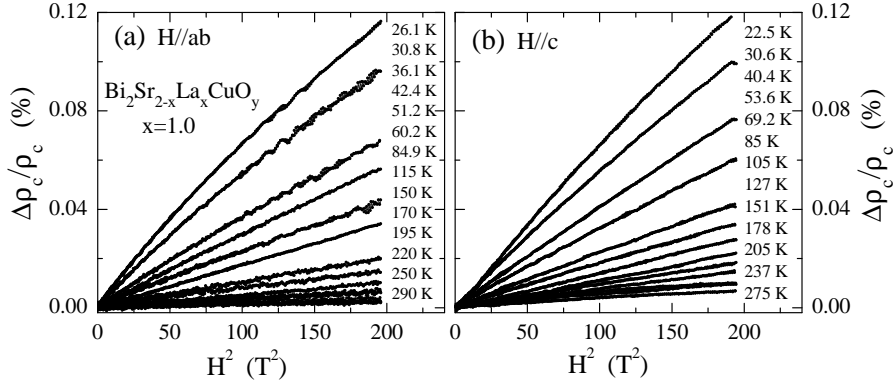}
\vspace{-5pt}
\caption{$H$ dependences of the transverse (a) and longitudinal (b)
$c$-axis MR for BSLCO with $x=1.0$.}
\label{fig3}
\end{figure}

As is mentioned in the introduction, whether the PG is affected by the
magnetic field or not depends crucially on its origin, and the MR data
are useful in clarifying this point. For all superconducting samples we
have observed that an application of the magnetic field causes a
considerable decrease in the normal-state $\rho_c$ in a rather wide
temperature range above $T_c$; as an example, precise MR data for
$x=0.5$ are shown in fig.~\ref{fig2}. Note that $\Delta\rho_c/\rho_c$ is
strictly proportional to $H^2$ at all temperatures down to $T_c$ when
$H$ is parallel to $ab$ [fig.~\ref{fig2}(a)], while the proportionality
to $H^2$ is observed only at high enough temperatures when $H$ is
parallel to $c$ [fig.~\ref{fig2}(b)]. In the non-SC ($x=1.0$) sample,
the MR turns out to be much weaker and positive at all temperatures
(fig.~\ref{fig3}), which gives evidence for an intimate connection
between the negative $c$-axis MR and superconductivity.

\begin{figure}[!b]
\onefigure[width=12.6cm]{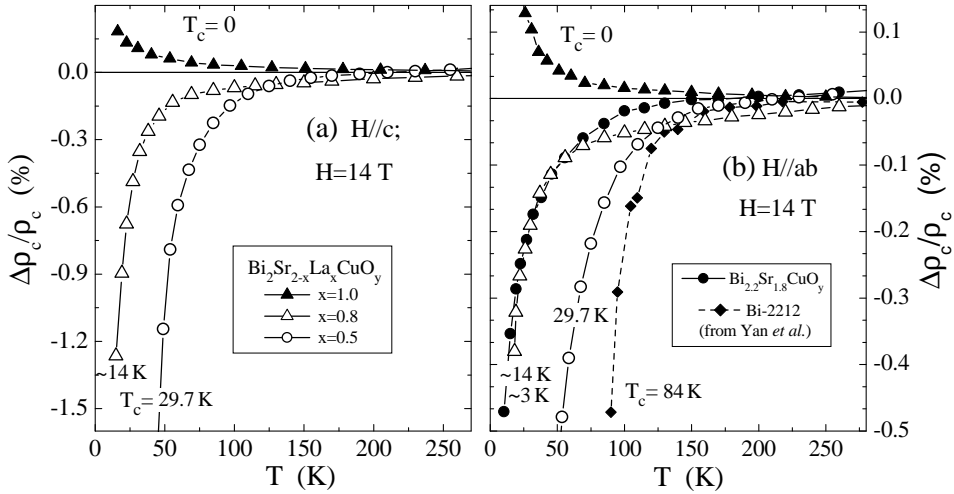}
\caption{$T$ dependences of the longitudinal (a) and transverse
(b) $c$-axis MR at $14\un{T}$ for BSLCO and BSCO crystals. The $T_c$
values are indicated near each curve. The MR data for Bi-2212 from Yan
\etal
\protect\cite{cMR} are shown for comparison.}
\label{fig4}
\end{figure}

The temperature dependences of the MR for underdoped and optimally-doped
crystals are depicted in fig.~\ref{fig4} (all the plotted data are from
the region where deviation from the $\Delta\rho_c/\rho_c \propto H^2$
behavior is insignificant). It is notable that a rapid growth of the
negative MR takes place with approaching $T_c$ that decreases with
decreasing hole density, rather than correlates with the upturn in
$\rho_c(T)$ which shifts to higher temperatures. Intriguingly, the
system seems to ``know'' whether it will eventually become
superconducting already at $\sim 150\un{K}$, where the sign of the MR
appears to be determined by the low-temperature ground state. This
indicates that the mechanism which causes the negative MR sets in
already at such high temperatures as $\sim 150\un{K}$ and that this
mechanism only exists when the sample becomes superconducting at low
temperatures.

\begin{figure}[t]
\onefigure[width=10.2cm]{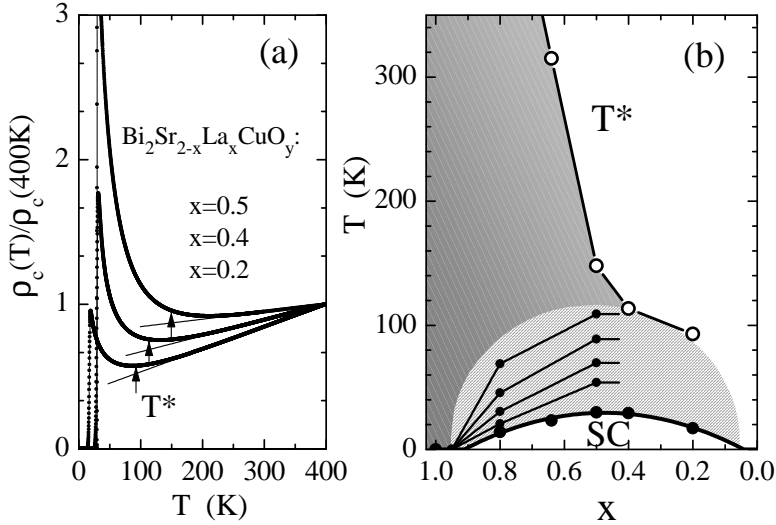}
\caption{(a) $c$-axis resistivity of Bi$_2$Sr$_{2-x}$La$_x$CuO$_y$.
$T^*$ is defined as a temperature where $\rho_c(T)$ deviates from its
linear extrapolation by 10\%. (b) Schematic phase diagram of BSLCO,
showing the PG region below $T^*$ (shaded) and the region where the
large negative MR is observed (hatched); a set of thin lines corresponds
to the longitudinal MR levels of (from above) 0.1, 0.2, 0.4, and 0.8\%.}
\label{fig5}
\end{figure}

Based on the behaviors of $\rho_c(T)$ and $\Delta\rho_c/\rho_c$, a
schematic phase diagram of BSLCO is sketched in fig.~\ref{fig5}, where
$T^*$ is defined as a temperature where $\rho_c(T)$ deviates from its
linear high-temperature behavior by 10\% [fig.~\ref{fig5}(a)]. The
evolution of $T^*$ with increasing hole density turns out to be notably
non-monotonic: a rapid change of $T^*$ in the underdoped region is
separated by a kink from a much slower decrease in the overdoped region.
On the other hand, the hatched area in fig.~\ref{fig5}(b), which marks
the region where considerable negative MR is observed, follows $T_c$
throughout the phase diagram, and apparently ignores the increase of
$T^*$ at low hole densities. For example, $|\Delta\rho_c/\rho_c|$ of the
$x=0.8$ sample (for which $T^*$ is above $400\un{K}$) stays less than
0.1\% at $14\un{T}$ in a wide temperature range below $T^*$ and starts
to grow rapidly only below $\sim 60\un{K}$, while in the $x=0.5$ sample
(for which $T^*$ is about $150\un{K}$) a rapid growth of the negative MR
is observed below $\sim 100\un{K}$; this trend is irrespective of the
field direction (fig.~\ref{fig4}). A simple estimate shows that away
from $T_c$ the magnetic-field scale for suppression of the upturn in
$\rho_c(T)$ (and thus of the normal-state PG) is extremely high: In the
$x=0.8$ sample, for instance, $\rho_c$ increases by $\sim$3 times upon
cooling from $T^*$ to $150\un{K}$, indicating that by $150\un{K}$,
$\sim$2/3 of the $c$-axis conductivity is wiped out by the PG. At the
same time, an application of $H=14\un{T}$ at $150\un{K}$ reduces
$\rho_c$ by only $\sim$0.04\%; therefore, even if the MR kept the steep
$\Delta\rho_c/\rho_c \propto H^2$ behavior up to the highest field, it
would require $\sim600\un{T}$ to completely suppress the PG. It is worth
noting that in YBa$_2$Cu$_3$O$_y$ the negative $c$-axis MR has also been
found to disappear when SC is killed by underdoping \cite{ourMR}.

The above result therefore distinguishes two essentially different
pseudogapped regions on the underdoped side of the phase diagram: one is
marked by the semiconducting $\rho_c(T)$ behavior and negligibly weak
magnetic-field dependence; the other is marked by the SC-related large
negative MR. This distinction can be understood if one assumes that
there are two distinct mechanisms that separately contribute to the PG
behavior. The predominant one that sets in around $T^*$ in underdoped
samples is almost insensitive to the magnetic field, gains strength with
decreasing hole concentration, and smoothly extends into the non-SC
region. The second one exists only in samples that show
superconductivity, causes further pseudogapping, and can be suppressed
by the magnetic field. On the overdoped side of the phase diagram, the
distinction between these two is not clear any more.

The first mechanism of the PG characterized by $T^*$ can probably be
associated with antiferromagnetic (AF) \cite{Batlogg} or stripe
\cite{Zachar} correlations that develop in the CuO$_2$ planes; these
correlations are fundamentally governed by the AF interactions and gain
strength as the hole density is reduced towards the AF region. Note that
the large exchange coupling $J$ ($\approx 0.12\un{eV}$) makes AF
interactions (and the PG caused thereby) robust against the magnetic
field. On the other hand, the second PG mechanism can most easily be
understood by attributing it to the SC fluctuations, which are
magnetic-field sensitive. One may naturally expect that the PG features
associated with SC fluctuations \cite{Varlamov} would occur in the
vicinity of the SC region, exhibit a singularity at $T_c$, and disappear
in non-SC compositions -- exactly the behavior that has been observed
for the secondary PG.

\begin{figure}[t]
\onefigure[width=10.0cm]{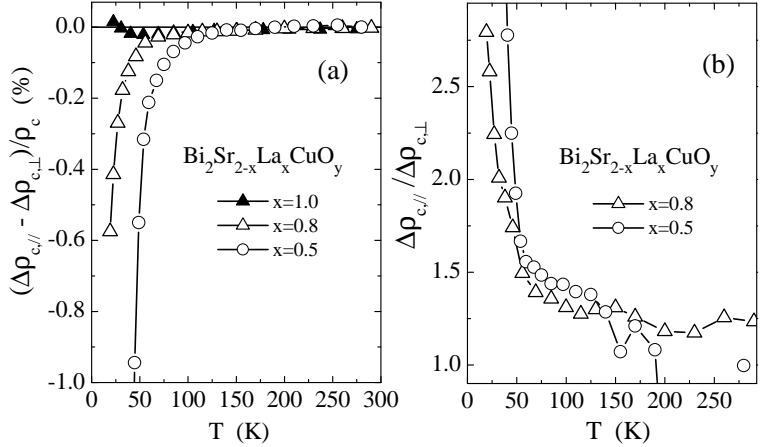}
\caption{(a) Difference between the longitudinal and transverse MR
obtained from the data in fig.~\ref{fig4}. (b) The MR anisotropy
$\Delta\rho_{c,\parallel}/ \Delta\rho_{c,\perp}$.}
\label{fig6}
\end{figure}

In fact, useful information on the PG mechanisms can be obtained from
the MR anisotropy shown in fig.~\ref{fig6}. At $T>140\un{K}$, the MR is
almost isotropic: difference between the longitudinal MR
($\Delta\rho_{c,\parallel}$ for $H\!\parallel\!c$) and transverse MR
($\Delta\rho_{c,\perp}$ for $H\!\perp\!c$) is virtually zero
[fig.~\ref{fig6}(a)] and thus the ratio $\Delta\rho_{c,\parallel}/
\Delta\rho_{c,\perp}$ is nearly one [fig.~\ref{fig6}(b)], which suggests
that the spin terms dominate at high temperatures. At lower
temperatures, however, the longitudinal MR significantly exceeds the
transverse MR in the SC samples (but not in the non-SC one); this means
that the magnetic field $H\!\parallel\!c$, which affects the {\it
in-plane} motion of carriers, becomes more effective in suppressing the
PG than $H\!\parallel\!ab$. This anisotropy may be readily understood if
the MR originates from the suppression of two-dimensional SC
fluctuations in CuO$_2$ planes \cite{Varlamov}, where the ``orbital''
effects of the magnetic field are relevant only for $H\!\parallel\!c$.
It is intriguing that the temperature range where the large negative MR
is observed appears to be notably broader than is usually expected for
SC fluctuations: in Bi$_{2.2}$Sr$_{1.8}$CuO$_y$, it extends up to $\sim
100\un{K}$, though $T_c$ is only $\sim 3\un{K}$. This corresponds nicely
to the recent scanning tunneling spectroscopy of the PG in overdoped
BSCO \cite{Kugler}, which also finds that the PG temperature is
extremely high compared to $T_c$.

The existence of two distinct mechanisms, which both cause the
depression of the DOS, might be the answer to the long-standing debates
on whether the PG has a SC \cite{STM,ARPES3,Varlamov} or non-SC
\cite{Tallon,Tallon2,NMR2} origin: there appears to be both a non-SC
pseudogap (that gains strength at low doping and tends to vanish in the
overdoped region \cite{Tallon2}) and a SC-induced suppression in the DOS
(that persists in a certain range above $T_c$ at all dopings
\cite{STM,Kugler}). The two mechanisms with contrasting doping and
magnetic-field dependences seem to be the main source of confusion in
studies of the PG in cuprates; measurements that probe the PG at high
temperatures and low doping apparently see the magnetic-field
insensitive part of the PG \cite{NMR3,NMR2}, while those performed at
low temperatures and higher doping see the SC-related part of the PG
that is sensitive to the magnetic field \cite{NMR1,overdoped}.

In summary, we present $c$-axis magnetotransport data of Bi-2201 that
strongly suggest existence of two distinct mechanisms for the pseudogap
in cuprates. The first one is almost insensitive to the magnetic field
and causes the primary pseudogap opening at $T^*$ in the underdoped
region; we discuss this may be related to the development of the
magnetic correlations or stripes. The second one is very magnetic-field
sensitive, which is the source of the large negative MR, and causes a
further pseudogapping as temperature is decreased towards $T_c$. This
second mechanism is found to be present only in superconducting samples
and is most strongly suppressed when the magnetic field is applied along
the $c$-axis, which points to its relation to the superconductivity.


\begin{thebibliography}{99}

\bibitem{review}
\Name{Timusk T. \and Statt B.}
\REVIEW{Rep. Prog. Phys.}{62}{1999}{61}.

\bibitem{STM}
\Name{Renner Ch. \etal}
\REVIEW{Phys. Rev. Lett.}{80}{1998}{149}.

\bibitem{ARPES3}
\Name{Ding H. \etal}
\REVIEW{Nature}{382}{1996}{51}.

\bibitem{ARPES4}
\Name{Norman M. R. \etal}
\REVIEW{Nature}{392}{1998}{157}.

\bibitem{ARPES1}
\Name{Ding H. \etal}
\REVIEW{Phys. Rev. Lett.}{78}{1997}{2628}.

\bibitem{2Gap1}
\Name{Krasnov V. M. \etal}
\REVIEW{Phys. Rev. Lett.}{84}{2000}{5860}.

\bibitem{2Gap2}
\Name{Suzuki M. \and Watanabe T.}
\REVIEW{Phys. Rev. Lett.}{85}{2000}{4787}.

\bibitem{Tallon}
\Name{Williams G. V. M., Tallon J. L. \and Loram J. W.}
\REVIEW{Phys. Rev. B}{58}{1998}{15053}.

\bibitem{Tallon2}
\Name{Tallon J. L. \and Loram J. W.}
\REVIEW{Physica C}{349}{2001}{53}.

\bibitem{NMR1}
\Name{Mitrovi\'{c} V. F. \etal}
\REVIEW{Phys. Rev. Lett.}{82}{1999}{2784}.

\bibitem{NMR3}
\Name{Gorny K. \etal}
\REVIEW{Phys. Rev. Lett.}{82}{1999}{177}.

\bibitem{NMR2}
\Name{Zheng G.-q. \etal}
\REVIEW{Phys. Rev. B}{60}{1999}{R9947}.

\bibitem{overdoped}
\Name{Zheng G.-q. \etal}
\REVIEW{Phys. Rev. Lett.}{85}{2000}{405}.

\bibitem{cMR}
\Name{Yan Y. F., Matl P., Harris J. M. \and Ong N. P.}
\REVIEW{Phys. Rev. B}{52}{1995}{R751}.

\bibitem{Rc_gap}
\Name{Matsuda A., Sugita S. \and Watanabe T.}
\REVIEW{Phys. Rev. B}{60}{1999}{1377};
\Name{Watanabe T., Fujii T. \and Matsuda A.}
\REVIEW{Phys. Rev. Lett.}{84}{2000}{5848}.

\bibitem{tunnel}
\Name{Suzuki M., Watanabe T. \and Matsuda A.}
\REVIEW{Phys. Rev. Lett.}{82}{1999}{5361}.

\bibitem{Ra}
\Name{Ito T., Takenaka K. \and Uchida S.}
\REVIEW{Phys. Rev. Lett.}{70}{1993}{3995}.

\bibitem{Bi_MR}
\Name{Heine G., Lang W., Wang X. L. \and Dou S. X.}
\REVIEW{Phys. Rev. B}{59}{1999}{11179}.

\bibitem{Varlamov}
\Name{Varlamov A. A., Balestrino G., Milani E. \and D. V. Livanov}
\REVIEW{Adv. Phys.}{48}{1999}{655}.

\bibitem{growth}
\Name{Ando Y. \and Murayama T.}
\REVIEW{Phys. Rev. B}{60}{1999}{R6991}.

\bibitem{Hanaki}
\Name{Ando Y. \etal}
\REVIEW{Phys. Rev. B}{61}{2000}{R14956}.

\bibitem{cont}
\Name{Lavrov A. N. \and Kozeeva L. P.}
\REVIEW{Physica C}{248}{1995}{365}.

\bibitem{ourMR}
\Name{Lavrov A. N., Ando Y., Segawa K. \and Takeya J.}
\REVIEW{Phys. Rev. Lett.}{83}{1999}{1419}.

\bibitem{Batlogg}
\Name{Batlogg B. \and Emery V. J.}
\REVIEW{Nature}{382}{1996}{20}.

\bibitem{Zachar}
\Name{Emery V. J., Kivelson S. A. \and Zachar O.}
\REVIEW{Phys. Rev. B}{56}{1997}{6120}.

\bibitem{Kugler}
\Name{Kugler M., Fisher \O., Renner Ch., Ono S. \and Ando Y.}
\REVIEW{Phys. Rev. Lett.}{86}{2001}{4911}.

\end{thebibliography}
\end{document}